\documentclass[a4paper]{jpconf}
\usepackage{graphicx}
\usepackage{amsmath}

\begin{document}
\title{GAN with an Auxiliary Regressor for the Fast Simulation of the Electromagnetic Calorimeter Response}

\author{Alexander Rogachev}

\address{HSE University}

\ead{airogachev@hse.ru}

\author{Fedor Ratnikov}

\address{HSE University, Yandex School of Data Analysis}

\ead{fedor.ratnikov@cern.ch}

\begin{abstract}
High energy physics experiments essentially rely on simulated data for physics analyses. However, running detailed simulation models requires a tremendous amount of computation resources. New approaches to speed up detector simulation are therefore needed. \\
The generation of calorimeter responses is often the most expensive component of the simulation chain for HEP experiments.
It was shown that deep learning techniques, especially Generative Adversarial Networks, may be used to reproduce the calorimeter response. However, those applications are challenging, as the generated responses need evaluation not only in terms of image consistency: different physics-based quality metrics should be also taken into consideration. \\
In our work, we develop a multitask GAN-based framework with the goal to speed up the response generation of the Electromagnetic Calorimeter (ECAL) of the LHCb detector at LHC. We introduce the Auxiliary Regressor as a second task to evaluate a proxy metric of the given input that is used by the Discriminator of the GAN. We show that this approach improves the stability of GAN and the model produces samples with better physics distributions. 
\end{abstract}

\section*{Introduction}
\label{intro}
The Large Hadron Collider (LHC) is the world’s largest collider built by the European Organization for Nuclear Research (CERN).  The LHCb experiment \cite{lhcb} at LHC focuses on studies of the heavy flavor physics, precise measurements of CP violation, and other effects in and beyond the Standard Model. The LHCb detector consists of several components, including an electromagnetic calorimeter (ECAL). The ability to simulate the expected detector response is a vital requirement for the physics analysis of the collected data and extracting physics results.
Currently, a baseline LHCb simulation based on the Geant4 package \cite{geant4} requires massive amounts of resources, being the most computationally expensive part of the experiments. In order to speed up the simulation, it is desirable to employ fast surrogate generative models like  Generative Adversarial Networks. However, the generated results should be consistent not only in terms of general quality metrics but rather in terms of physics metrics. 

\section{Dataset}
\label{data}
\begin{figure}[htb]
        \begin{center}
        \includegraphics[ width=0.9\linewidth]{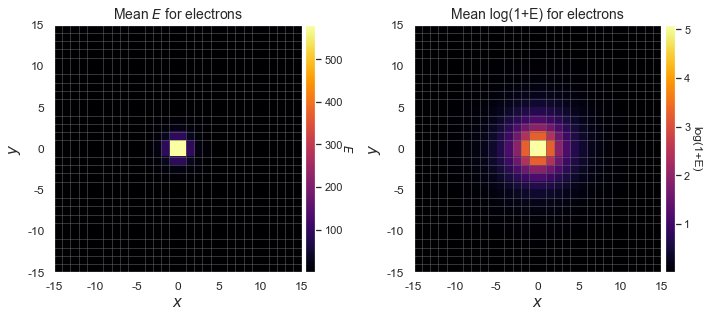}
        \end{center}
        \caption{\label{fig:ecal} Visualization of the ECAL response dataset.}
    \end{figure}

The dataset that we used through the experiments contains information about the interactions of electrons inside the electromagnetic calorimeter. ECAL uses "shashlik" technology of alternating layers of lead and scintillation plates. Readout cells in different modules are $4\times 4$, $6\times 6$, and $12\times 12$ $cm^2$ in size, thus it is possible to obtain response for all granularities by aggregating responses of $2\times 2$ $cm^2$ logical cells. All the events in the data correspond to electrons with a given momentum and direction entering the calorimeter at a given location.
Such an electron produces an electromagnetic shower in the ECAL. All energies deposited in the scintillator layers of one cell are summed up and produce a matrix of the energies corresponding to the ECAL response for such impact electron, as presented in Figure \ref{fig:ecal}. Dataset of such ECAL responses in $30\times 30$ cells matrix of $2\times 2$ $cm^2$ cells approximately centered on the energy cluster location was produced using GEANT4 package. 

\section{GANs for electromagnetic calorimeter response simulation}
\label{CaloGan}
Generative adversarial network (GAN) is a widely used approach to building a generative model. GAN consists of two components, generator and discriminator, which compete against each other to reproduce objects from a given distribution \cite{gan}. 
The generator (G) learns to map some easy-to-draw distribution, e.g. a standard normal distribution, to the target distribution. The feedback from discriminator (D), is used to improve the generator. The main goal of the discriminator is to find the discrepancy between the reference real objects and those produced by the generator ones.

More formally, this minimax game can be represented as follows:
    \begin{equation} \label{eq2}
    \begin{split}
        \min_G\max_D \mathrm{E_{x \sim p_{data}(x)}}[logD(x)]+\mathrm{E_{z\sim N(0,I)}}[log(1-D(G(z)))],
    \end{split}
    \end{equation}
    
    \noindent where $p_{data}$ is a true data distribution, $D(x)$ is the output of the discriminator.

The idea of using Generative Adversarial Networks to perform the simulation in High Energy Physics was proposed by Paganini et al \cite{calogan}, it was further developed in \cite{Chekalina} with the application of Wasserstein CGAN. GANs performance was also compared with the model based on the idea of Conditional Variational Autoencoder and their combination, considering CGAN to provide the best results \cite{Sergeev_2021}. 

In our previous work \cite{Rogachev_Ratnikor_2021}, we proposed to improve the quality of the generative model by adding Self-Attention layers to the previous best-performing architecture, thus allowing CNN-based networks to catch and use long-range relationships between image regions through the training process. The training process was also stabilized by introducing Spectral Normalization \cite{SN} into both models. 

To evaluate the quality of generated samples and compare the performance of the models, we use PRD-AUC \cite{prdauc}, precision-recall-style metrics for the distribution case. It disentangles the quality of generated samples (precision) from the proportion of target distribution, covered by the generated distribution (recall):
    
    \begin{equation} \label{prd_eq}
        \mathrm{PRD(Q,P)} = \{(\theta \alpha(\lambda),\theta \beta(\lambda)) | \lambda \in (0,\infty),\theta \in [0,1]\},
    \end{equation}    
    
    \noindent where P and Q are distributions, that are defined on a finite state space,
    
    $$\alpha(\lambda) = \sum_{\omega \in \Omega}\min(\lambda P(\omega),Q(\omega)),$$
    
    $$\beta(\lambda) = \sum_{\omega \in \Omega}\min(P(\omega), \frac{Q(\omega)}{\lambda}).$$
    
 In order to evaluate images not only in terms of general quality but in terms of physics metrics as well, we use the minimum of two PRD-AUC scores, evaluated over raw images and an arbitrary selected illustrative set of the following physics statistics:
\
    \begin{itemize}
        \item shower asymmetry along and across the direction of inclination;
        \item shower width ;
        \item the number of cells with energies above a certain threshold, the sparsity level.
    \end{itemize}
    
This approach works properly with discrete distributions only, thus we unite the points from real and generated distributions and then cluster them using MiniBatchKMeans. Then we take the pair of histograms of discrete distributions over the cluster centers and evaluate the PRD. To compare models during our experiments, we use 400 clusters.

\section{Auxiliary Regressor}

In order to improve the quality of produced distributions, especially of those statistics that we use during quality evaluation, we propose the following extension of the Discriminator.

We call it Auxiliary Regressor and its goal is to evaluate some particular metrics that we want to reproduce. It shares the first layers with the regular discriminator, and we train both networks simultaneously.   
The intuition here is that by introducing an additional task into the training procedure we let our model catch some general information that can be useful for both objectives. Moreover, we try to provide the NN with the information about the desired metrics, expecting it to learn it as now discriminators can detect generated objects with badly reproduces metrics values. Thus, it even becomes possible to optimize a quality metric that was not differentiable before, as now we can use backpropagation to train a network that approximates it. 

\begin{figure}[htb]
        \begin{center}
        \includegraphics[ width=0.9\linewidth]{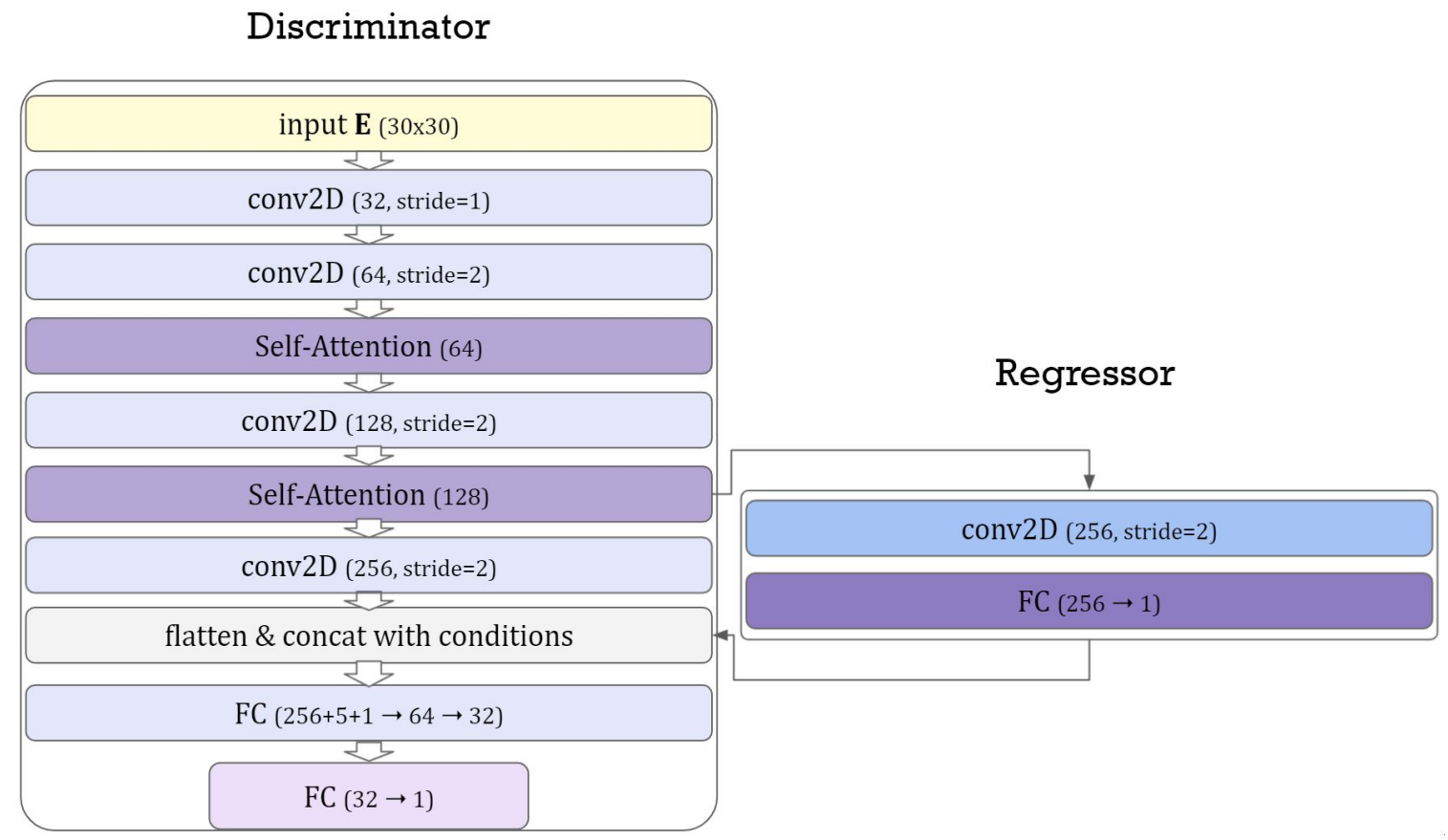}
        \end{center}
        \caption{\label{fig:arch} Example of the Discriminator with the AUX-Regressor}
    \end{figure}

The general architecture is presented in Figure \ref{fig:arch}. The output of the last SA layer comes to the Regressor part as well as to another convolutional layer inside the default Discriminator. The output of the Regressor is used to calculate regression loss (we use MAE in our case) and it is passed back to the Discriminator where we stack it with other conditions.

In our experiments, we added the regressor to models with different architectures: CNN and the same model with two Self-Attention Layers (the best model from \cite{Rogachev_Ratnikor_2021}). We also evaluate the impact of the Spectral Normalization that we studied in our previous work. The final loss that we use to train models with Auxiliary Regressor consists of two parts: adversarial hinge loss and mean absolute error, related to the regressor. 

\section{Results and discussion}

The results of our experiments are presented in Table \ref{tab:results}. In all cases models with the additional regression loss perform better than the corresponding models without it. The Spectral Normalization allows to slightly stabilize the training procedure and improve the quality even further.

\begin{table}[h]
\caption{\label{tab:results}Quality of trained models.}
\begin{center}
\begin{tabular}{llll}
\br
Architecture &Raw PRD&Phys PRD\\
\mr
CNN baseline &0.669&0.877\\
CNN + Regressor &0.822&0.950\\
CNN + Regressor + SN&0.908&0.957\\
CNN + SA &0.989&0.983\\
CNN + Regressor + SA + SN &0.994&0.992\\
\br
\end{tabular}
\end{center}
\end{table}

\begin{figure}[htb]
        \begin{center}
        \includegraphics[ width=0.9\linewidth]{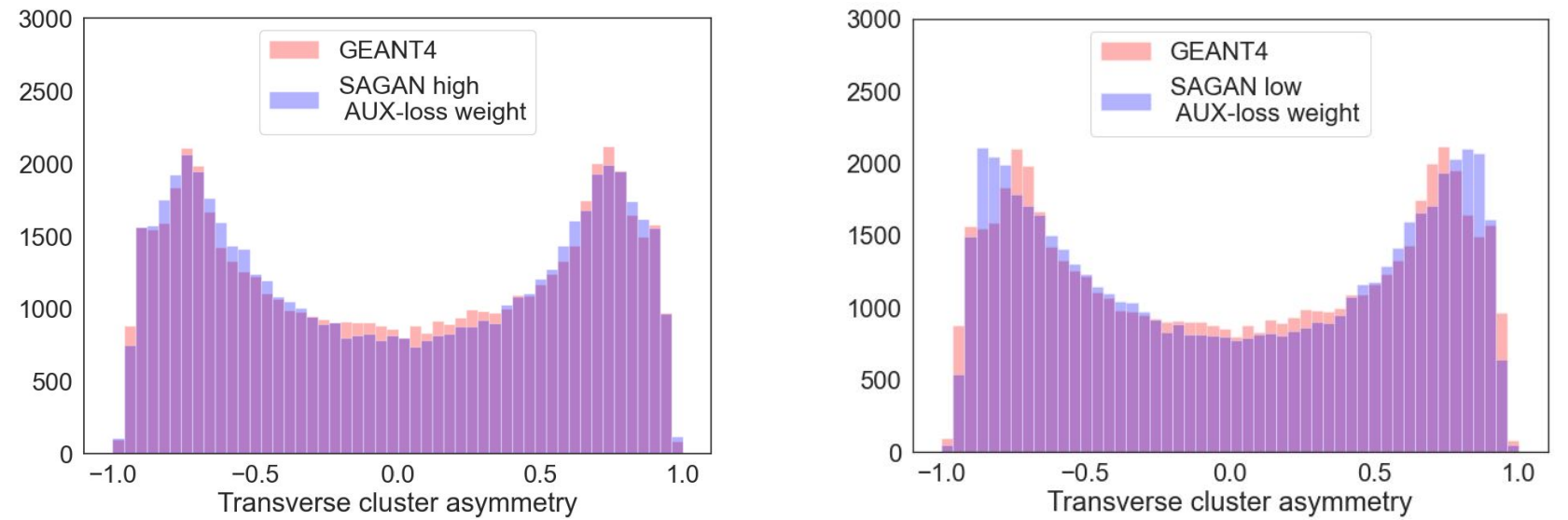}
        \end{center}
        \caption{\label{fig:assym} Asymmetry distribution of models with different regression loss weights}
    \end{figure}

We also studied how the way of choosing weights of losses affects the quality of a chosen metric to reproduce the distribution. In case we use relatively high weight, the peaks in the reference physics distribution (see Figure \ref{fig:assym}) of the metric distribution that was evaluated using generated objects become closer to the original ones, and it resolves the problem that was indicated in our previous studies. As we decrease the weight, the asymmetry distribution becomes similar to the one that the model with no additional regressor, as may be expected.

\section*{Conclusion}

In this paper, we propose to train an instance of Generative Adversarial Network in multitask manner by extending GAN discriminator by an additional auxiliary regressor that evaluates one of the metrics over generated objects that we want to reproduce. This may be an efficient technique for better reproducing physics characteristics of the generative models, which is demonstrated on the use case of the fast simulation of the electromagnetic calorimeter for the LHCb experiment. This approach allows us to boost the quality of generated distribution in terms of PRD, as the distribution of the evaluated metric, that was used as the target of the regressor, became closer to the original one. 

Future work may be focused on the application of the proposed auxiliary regression in the case of other datasets where some particular metrics or qualities of generated objects should be reproduced. This approach requires the metric to be computable on the object level, but it is not necessary for this function to be differentiable, as introducing the regressor as a neural network makes it differentiable. 

\section*{Acknowledgments}

The study was implemented in the framework of the Basic Research Program at the National Research University Higher School of Economics (HSE University) in 2022.

\section*{References}

\bibliography{main}

\end{document}